\documentclass[reprint,amsmath,amssymb,aps,prx,groupedaddress,nofootinbib,superscriptaddress]{revtex4-2}
\usepackage{setspace}
\usepackage[latin1]{inputenc}
\usepackage{graphicx}
\usepackage{physics}
\usepackage{amssymb,amsmath,braket,mathdots}
\usepackage{upgreek}
\usepackage{bm,bbold}
\usepackage[dvipsnames]{xcolor}
\usepackage{graphicx,tikz}
\usepackage{epstopdf,psfrag}
\usepackage{relsize,amsbsy}
\usepackage[export]{adjustbox}
\usepackage{graphicx,xcolor,tikz}
\usepackage{subfigure}
\usepackage[normalem]{ulem}
\usepackage{color, colortbl}
\usepackage{mathbbol}
\usepackage{bbm}

\usepackage[colorlinks,linktocpage]{hyperref}
\hypersetup{
	citecolor  = NavyBlue,
	linkcolor  = NavyBlue,
	urlcolor = NavyBlue
}

\DeclareMathOperator{\cTr}{\mathbb{Tr}}

\newcommand{\iu}{\mathrm{i}} 
\newcommand{\eu}{\mathrm{e}}

\newcommand{\be}{\begin{equation}}
	\newcommand{\ee}{\end{equation}}
\newcommand{\ba}{\begin{align}}
	\newcommand{\ea}{\end{align}}
\newcommand{\bit}{\begin{enumerate}}
	\newcommand{\eit}{\end{enumerate}}

\newcommand*\diff{\mathop{}\!\mathrm{d}}
\newcommand*\ddiff[1]{\mathop{}\!\mathrm{d}#1\,}
\newcommand*\Diff{\mathop{}\!\mathcal{D}}
\newcommand{\non}{\nonumber}
\newcommand{\dg}{\dagger}
\newcommand{\ve}[1]{{\bf #1}}

\newcommand{\ch}{\mathrm{q}}

\newcommand{\msf}[1]{\mathsf{#1}}

\newcommand{\cket}[1]{| #1 )}
\newcommand{\cbra}[1]{( #1 |}

\newcommand{\mband}{\rm m}
\newcommand{\lband}{\rm l}
\newcommand{\mG}{\mathsf{G}}
\newcommand{\mSigma}{\mathsf{\Sigma}}
\newcommand{\mX}{\mathsf{X}}
\newcommand{\mv}{\mathsf{v}}
\newcommand{\mV}{\mathsf{V}}

\definecolor{bananayellow}{rgb}{1.0, 0.88, 0.21}
\definecolor{straw}{rgb}{0.32, 0.28, 0.1}

\begin{document}
	\title{A Quantum Many-Body Approach for Orbital Magnetism in Correlated Multiband Electron Systems}
	\author{Mengxing Ye}
	\affiliation{Department of Physics and Astronomy, University of Utah, Salt Lake City, UT, 84112, USA}
	\date{\today} 

	\begin{abstract}
		Orbital magnetism is a purely quantum phenomenon that reflects intrinsic electronic properties of solids, yet its microscopic description in interacting multiband systems remains incomplete. We develop a general quantum many-body framework for orbital magnetic responses based on the Luttinger--Ward functional. Starting from the Dyson equation, we reformulate the thermodynamic potential in a weak magnetic field and construct a controlled expansion in powers of $B$ applicable to correlated electron systems. A key technical advance is a modified ``Fourier'' representation using noncommutative coordinates, which allows the thermodynamic potential to be expressed in an effective momentum space where the magnetic field acts perturbatively. This formulation makes analytic progress possible within the Moyal algebra. As an application, we derive the spontaneous orbital magnetization for approximations whose skeleton diagrams contain no fermion loop enclosing magnetic flux (e.g., Hartree--Fock, self-consistent RPA, and single-site dynamical mean-field theory). Vertex corrections consistent with the self-energy are included automatically, without a separate diagrammatic analysis, and the result is expressed entirely in terms of the zero-field Hamiltonian renormalized by the self-energy. For frequency-dependent but Hermitian self-energies, we generalize the orbital magnetic moment and Berry curvature to momentum-frequency space and identify two gauge-invariant contributions built from these quantities. For frequency-independent self-energies the result reduces to the familiar geometric formula for noninteracting systems. This framework provides a unified foundation for computing orbital magnetic responses in correlated multiband materials.
	\end{abstract}
	\maketitle

	\tableofcontents

	\section{Introduction}

	Recent advances in highly tunable two-dimensional heterostructures, such as moir\'e materials, have created a fertile platform for exploring correlated multiband electron systems. In the noninteracting limit, the Bloch wave functions of multiband Hamiltonians endow the Brillouin zone with a rich geometrical structure, often termed \emph{quantum geometry}~\cite{Provost1980,Resta2011}. Establishing quantitative connections between quantum-geometric quantities and experimentally measurable observables -- especially in the presence of strong electronic correlations -- has become an active frontier of research~\cite{QGReview2023,jiabin_2024,Gao2025quantum}.

	A direct and informative probe of quantum materials is the application of an external magnetic field. Electrons couple to the field through both spin and orbital channels; the orbital coupling introduces Peierls phases that can alter the spectrum and wave functions in nonperturbative ways. In strong fields, this coupling can produce quantum Hall phases~\cite{Klitzing1980,Tsui1982}, while in weak fields it leads to oscillatory contributions to the thermodynamic potential (e.g., de Haas--van Alphen effect) with period scaling as $1/B$ because of Landau quantization~\cite{Shoenberg1984,Abrikosov2017}.

	Beyond these oscillations, orbital magnetic responses also contain smooth contributions analytic in powers of $B$. In systems that spontaneously break time-reversal symmetry, a finite orbital magnetization can appear even without an external field. The spontaneous orbital magnetization, \( M_{\mathrm{orb}} \), couples linearly to a weak external field \( B \); the coefficient of the $B$-linear term in the thermodynamic potential defines \( M_{\mathrm{orb}} \). For noninteracting systems this quantity is well understood~\cite{xiao_shi_niu_2005,Xiao_Chang_Niu_RMP,VanderbiltMorb2005,CeresoliPRB2006,Thonhauser2011OrbitalMagnetization}: it comprises two gauge-invariant geometric contributions, one from the intrinsic orbital moment of Bloch electrons and another from the Berry curvature of occupied states, as captured by the St\v{r}eda relation~\cite{Streda1982,assa-2019}, which links density changes with magnetic field to the Hall conductivity. Recent observations of orbital ferromagnetism in moir\'e systems~\cite{Sharpe2019Emergent,Serlin2019IntrinsicQAH,Lu2019SuperconductorsOrbital,Chen2020TunableChern,Song2021DirectVisualization,He2021CompetingCorrelated,Zeng2023ThermodynamicFractionalChern,Lu2023FractionalQuantum,Polshyn2020ElectricalSwitching,Redekop2024DirectMagneticImaging} motivate a quantitative theory of orbital magnetization in strongly correlated materials, where interactions can qualitatively modify magnetic responses. 

	Extending these noninteracting results to interacting systems is a fundamental theoretical challenge. The modern theory of orbital magnetization (developed in the 2000s) relied on semiclassical wave-packet methods~\cite{xiao_shi_niu_2005,Xiao_Chang_Niu_RMP} and Wannier-based formulations~\cite{VanderbiltMorb2005,CeresoliPRB2006,Thonhauser2011OrbitalMagnetization} for noninteracting systems. Later works reproduced these results using spatially varying magnetic fields~\cite{NiuMorbQ2007} or EM-gauge-invariant~\footnote{For clarity, we use ``EM-gauge" for the electromagnetic gauge transformation defined in Eq.~\eqref{eq:EMgauge}, and use ``gauge" for gauge transformation to the multiband Bloch states as defined in Eq.~\eqref{eq:gauge}.} Green's functions~\cite{PALee2011PRB,RaouxPRB2015,Haldane2021GaugeInvariant}. Recently, several studies~\cite{Kang2025Orbital,Liu2025Orbital,Zhu2025Magnetic} extended these ideas to interacting systems within Hartree--Fock or mean-field approximations. However, a fully systematic method to compute orbital magnetic responses in general interacting systems is still lacking.

	A natural route is to compute the thermodynamic potential directly -- not just a mean-field energy -- in a weak magnetic field. Recent mean-field works evaluate an energy functional~\cite{Kang2025Orbital,Liu2025Orbital,Zhu2025Magnetic}, which is harder to generalize beyond mean-field. Another conceptual difficulty is separating oscillatory contributions from smooth, perturbative terms~\cite{ye_wang_first,Wang2025Bosonized}. Earlier approaches introduced a spatially varying magnetic field with momentum $q$~\cite{NiuMorbQ2007,PinesNozieres1966,Vignale_1988,GiulianiVignale2005,Tremblay_2014,Chubukov_dia_2023}, leading to expansions in powers of $(q\ell_B)^{-1}$, where $\ell_B$ is the magnetic length. Practically, orbital magnetic responses are defined in the regime where quantum oscillations are exponentially suppressed, $\sim \eu^{-T/\omega_c}$, so one needs a controlled expansion in the dimensionless parameter $\omega_c/T$ (with $\omega_c$ the cyclotron frequency and $T$ the temperature). Constructing such an expansion is essential to define smooth coefficients in powers of $B$ while suppressing oscillatory effects.

	The goal of this work is to formulate such an expansion for the thermodynamic potential of \emph{interacting multiband systems}. Because the magnetic field couples via the gauge potential, ordinary translational symmetry is broken and the conventional Fourier transform is not a convenient basis. Magnetic translation symmetry remains, but its unit cell becomes large in the weak-field limit, which obscures analytic tractability. A central development of this paper is the identification of a consistent scheme of modified ``Fourier'' transforms -- unitary and exact -- under which the magnetic field acts \emph{locally} in an effective momentum space. This construction enables a systematic expansion of the thermodynamic potential in powers of $B$ written in terms of the full Green's function and self-energy as self-consistent solutions to the Dyson equations. Because the thermodynamic potential is evaluated with the self-consistent Green's function and self-energy in the field, vertex corrections consistent with the self-energy are included automatically, while in response-function approaches they must be constructed diagrammatically to preserve EM-gauge invariance~\cite{BaymKadanoff1961,Baym1962}.

	As an application, we derive the spontaneous orbital magnetization \( M_{\mathrm{orb}} \) for interacting systems, showing that, when no fermion loop of the two-particle-irreducible (2PI) functional encloses magnetic flux [e.g., Hartree--Fock, self-consistent RPA, and single-site dynamical mean-field theory (DMFT)], it can be expressed entirely in terms of the zero-field renormalized Hamiltonian, the sum of the bare Bloch Hamiltonian and the self-energy evaluated at $B=0$. For frequency-dependent but Hermitian self-energies, we generalize the orbital magnetic moment and Berry curvature to momentum-frequency space and identify two gauge-invariant contributions built from these quantities. For frequency-independent self-energies the result reduces to the familiar noninteracting form with a self-energy--renormalized Hamiltonian. This establishes a unified formalism to compute orbital magnetic responses of correlated electrons from zero-field quantities.

	The remainder of the paper is organized as follows:
	Section~\ref{sec:review} formulates the problem and summarizes the main theoretical results.
	Section~\ref{sec:DysonMorb} reviews the EM-gauge-covariant structure of Green's functions and the Dyson equation in a magnetic field using the Moyal algebra.
	Section~\ref{sec:Omega}, the key development of this paper, presents a systematic expansion of the thermodynamic potential, treating the trace-log term and interaction contributions.
	Section~\ref{sec:Omega3} applies the formalism to compute the spontaneous orbital magnetization, generalizes the geometric quantities in momentum-frequency space and discusses its reduction to the familiar geometric formula for static self-energies.
	Section~\ref{sec:summary_outlook} summarizes and discusses the main results and outlines future directions.
	For clarity, notation is summarized in Appendix~\ref{app:notation}. The remaining appendices contain detailed derivations that supplement the main text.

	\paragraph*{Notes added.}
	Upon completion of this work, we became aware of several recent works~\cite{Kang2025Orbital,Liu2025Orbital,Zhu2025Magnetic} that study spontaneous orbital magnetization in multiband systems within Hartree--Fock or related mean-field approaches. In the limit of static self-energies, our results are compatible with these studies. The present work focuses on a general many-body formulation based on the Luttinger--Ward functional and does not rely on a specific mean-field approximation.

	\section{Formulation and Summary of Main Results}
	\label{sec:review}

	Before presenting detailed derivations, we first formulate the problem, summarize the theoretical approach and highlight main results.

	We consider an interacting multi-orbital system in a magnetic field $B$. For concreteness we focus on a two-dimensional system with the magnetic field perpendicular to the plane and neglect spin and Zeeman coupling. The noninteracting Hamiltonian is introduced via the Peierls substitution,
	\begin{align}
		\mathcal{H}_g
		&= -\sum_{\langle ij\rangle}\sum_{\alpha\beta} 
		t_{ij,\alpha\beta}\,
		a^{\dagger}_{i,\alpha}\,
		\eu^{\iu \ch \int_{r_{j,\beta}}^{r_{i,\alpha}} A(r')\cdot \diff r'}\,
		a_{j,\beta},
		\label{eq:H0_1}
	\end{align}
	where $i,j$ denote lattice sites (unit cells), $\alpha,\beta$ label orbitals, and $r_{i,\alpha}$ is the position of orbital $\alpha$ in unit cell $i$; the Wilson line follows the straight path between the two orbital positions. Grouping the hoppings by the bond vector $\delta=r_{i,\alpha}-r_{j,\beta}$, the hopping amplitudes form a matrix $\mathsf{t}_{\delta}$ in orbital space.\footnote{Throughout, we use \textsf{sans serif} font for matrices in orbital/band space.} 
	The charge $\ch$ is kept general; for electrons, $\ch=-e$. 

	For weak magnetic field such that the magnetic length is much larger than the lattice constant, i.e.\ $\abs{\ch B} a^2 \ll 1$, discrete lattice translations are not crucial~\cite{ye_wang_second}. The Hamiltonian can be rewritten in terms of fermion fields in continuous space via two Fourier transforms:
	\begin{align}
		\mathcal{H}_g = \int \diff^2 r \sum_{\delta} 
		a^{\dagger}_r 
		\left( -\mathsf{t}_{\delta}\, 
		\eu^{\iu \ch \int_{r-\delta}^{r} A(r')\cdot \diff r'} \right)
		a_{r-\delta}.
	\end{align}

	The interacting thermodynamic potential follows from the Luttinger--Ward functional,
	\begin{align}
		\Omega_{\rm int}
		= -T\, \Tr \ln \mG^{-1}
		- T\, \Tr (\mSigma \circ \mG)
		+ \Phi(\mG),
	\end{align}
	where $\Phi(\mG)$ is the sum of all two-particle-irreducible (skeleton) vacuum diagrams built from $\mG$ (App.~\ref{app:LWintroduction}), and $\mG$ and $\mSigma$ satisfy the Dyson equations,
	\begin{subequations}
		\begin{align}
			\left(\mG_0^{-1} - \mSigma\right)_{12}
			&= \mG^{-1}_{12}, \label{eq:Dyson1}\\
			\frac{\delta \Phi(\mG)}{\delta \mG_{12}} 
			&= T\,\mSigma_{21}. \label{eq:Dyson2}
		\end{align}
	\end{subequations}
	Here indices $l = 1,2,\ldots$ denote space--imaginary-time coordinates $(r_l,\tau_l)$, and the bilocal convolution is
	\begin{align}
		(f\circ g)_{1 1'}
		= \int \ddiff{2} f_{12}\, g_{2 1'}, \quad
		\ddiff{l} = \diff\tau_l\, \diff^2 r_l.
	\end{align}

	The noninteracting kernel $\mG_0^{-1}$ has an EM-gauge-covariant form,
	\begin{align}
		&(\mG_0^{-1})_{12}
		= \eu^{\iu\ch \int_{r_2}^{r_1} A(r')\cdot \diff r'}\,
		\delta(\tau_1-\tau_2)\times\non\\
		&  \quad
		\left[
		\delta(r_1-r_2)(-\partial_{\tau_2}+\mu)
		- \sum_{\delta}
		\delta(r_1-\delta-r_2) (-\mathsf{t}_{\delta})
		\right],
		\label{eq:G0inv}
	\end{align}
	i.e., a Wilson line multiplied by a translationally invariant kernel. As argued in Sec.~\ref{sec:DysonMorb}, EM-gauge covariance and the variational relation~\eqref{eq:Dyson2} imply that, for density--density interactions, $\mG_{12}$ and $\mSigma_{12}$ share the same structure. For brevity we write any bilocal function $\mathcal{F}$ as~\cite{Luttinger1961,BychkovGorkov1962}
	\begin{align}
		\mathcal{F}(\tau_1, r_1; \tau_2, r_2)
		= \eu^{\iu\ch \int_{r_2}^{r_1} A(r')\cdot \diff r'}\,
		\mathcal{F}'(\tau_1-\tau_2, r_1 - r_2).
		\label{eq:bilocal}
	\end{align}

	Using this property, the first Dyson equation can be written in momentum space as
	\begin{align}
		\Big[
		(\iu\omega_n - \tfrac{1}{2}\partial_T + \mu)
		- \mathsf{H}_{p,n}\Big]\star
		\mG'(p,\omega_n)= 1,
		\label{eq:DysonFinal}
	\end{align}
	where $\mathsf{H}_{p,n} = \mathsf{h}_p + \mSigma'(p,\omega_n)$ is the renormalized Hamiltonian, $\mathsf{h}_p = \sum_\delta (-\mathsf{t}_\delta)\, \eu^{-\iu p\cdot \delta}$ is the zero-field Bloch Hamiltonian, and $\star$ denotes the Moyal product,
	\begin{align}
		\star = \eu^{\frac{\iu\theta_{ab}}{2}\overleftarrow{\partial}_{p_a}
			\overrightarrow{\partial}_{p_b}}, \quad
		\theta_{ab} = \ch B \epsilon_{ab}.
	\end{align}
	This structure reflects the noncommutativity of mechanical momentum, $[\hat{p}_a, \hat{p}_b]=\iu\theta_{ab}$. In the limit $B\rightarrow 0$, the equation reduces to the standard Dyson equation. Both $\mG'$ and $\mSigma'$ depend on $B$; in the regime $\omega_c/T \ll 1$ they can be solved order by order using this formulation. This Dyson equation appears in earlier literature for noninteracting~\cite{PALee2011PRB} and interacting~\cite{Finkelstein2003,KitaArai2005InteractingBloch,Goldman2023} cases. Similar results have also been discussed in~\cite{Onoda2006NonEquilibrium}.

	While the Dyson equation admits a systematic expansion in $B$, computing the thermodynamic potential -- in particular the $-T\, \Tr \ln \mG^{-1}$ term -- is nontrivial because of the Wilson line and the logarithm. To address this, we reformulate the problem in the fermionic path integral and introduce an invertible transformation whose momentum label corresponds to the first-quantized mechanical momentum. Using noncommutative-geometry techniques~\cite{Szabo2003,Douglas2001rmp} we show that the path integral can be expressed entirely in terms of $\mG'$ and $\mSigma'$ in an effective momentum space governed by the Moyal algebra [see Eq.~\eqref{eq:OmegaA2}]. In that formulation the $B$ dependence is captured both in the algebra and via the $B$ dependence of $\mG'$ and $\mSigma'$ determined by the Dyson equation. The action resembles that of a noncommutative field theory~\cite{Douglas2001rmp,Szabo2003}, although here the noncommutative coordinate is momentum rather than position. To our knowledge this is the first work to demonstrate this analogy for electronic systems in a weak magnetic field.

	Expanding the path integral in powers of $B$ requires careful handling of open derivatives in the Moyal algebra, which is especially subtle in the presence of the logarithm. We develop a systematic procedure to expand the thermodynamic potential [see Eqs.~\eqref{eq:OmegaA3} and~\eqref{eq:OmegaB}], and apply it to compute the spontaneous orbital magnetization,
	\begin{align}
		M_{\rm orb}
		= -\frac{1}{\mathcal{V}}
		\left.\frac{\delta\Omega_{\rm int}}{\delta B}\right|_{\mu, T, B=0}.
		\label{eq:Morb_def}
	\end{align}
	We find that, for the class of approximations specified in Sec.~\ref{sec:Omega2}, $M_{\rm orb}$ can be expressed entirely in terms of the zero-field Green's function $\mG'^{(0)}(p,\omega_n)$ and generalized velocity operators $\mathsf{v}^{(0)}_{a}=\partial_{p_a}\mathsf{H}^{(0)}_{p,n}$ [see Eq.~\eqref{eq:Morb_v}], where $\mathsf{H}^{(0)}_{p,n} = \mathsf{h}_p + \mSigma'^{(0)}(p,\omega_n)$ and the superscript $(0)$ denotes zero-field quantities. For Hermitian $\mathsf{H}^{(0)}_{p,n}$, its eigenvectors define generalized Bloch wave functions in momentum-frequency space. We find that $M_{\rm orb}$ consists of two gauge-invariant contributions [see Eq.~\eqref{eq:Morb_hermitian}], one from a generalized magnetic moment [see Eq.~\eqref{eq:orbmoment}] and another from a generalized Berry curvature [see Eq.~\eqref{eq:berrycurvature}]. When the self-energies are frequency-independent, the result reduces to the familiar noninteracting form with a self-energy--renormalized Hamiltonian.

	\section{Dyson Equation in a Magnetic Field}
	\label{sec:DysonMorb}

	We now review the derivation of the Dyson equation in the presence of a magnetic field. While this material is known~\cite{Luttinger1961,BychkovGorkov1962,PALee2011PRB,Finkelstein2003,Goldman2023}, we present a detailed discussion that clarifies the gauge-covariant structure and the physical implications of the Moyal algebra.

	\paragraph*{EM-Gauge-covariant structure.}
	The full Green's function $\mG$ and self-energy $\mSigma$ must be EM-gauge covariant and can therefore be written as a product of a Wilson line and a translationally invariant function, denoted $\mG'$ and $\mSigma'$ (see Eq.~\eqref{eq:bilocal}). To see this, note that under an EM-gauge transformation 
	\begin{align}
		A(r)\rightarrow A(r)+\nabla \Lambda(r),
		\label{eq:EMgauge}
	\end{align}
	 the fermion operator transforms as $a_r \rightarrow \eu^{\iu \ch \Lambda(r)} a_r$. Consequently, the Green's function transforms as
	\[
	\mG(r,r') \rightarrow \eu^{\iu \ch \Lambda(r)} \mG(r,r') \eu^{-\iu \ch \Lambda(r')},
	\]
	which matches the transformation of the Wilson line
	\[
	\eu^{\iu \ch \int_{r'}^{r} A(r'')\cdot \diff r''} \rightarrow \eu^{\iu \ch \Lambda(r)} \eu^{\iu \ch \int_{r'}^{r} A(r'')\cdot \diff r''} \eu^{-\iu \ch \Lambda(r')}.
	\]
	Thus $\mG(r,r')$ can be written as a Wilson line times an EM-gauge-invariant bilocal function depending only on $r-r'$. The same argument applies to $\mSigma$ for density--density interactions -- from Eq.~\eqref{eq:Dyson2}, $\mSigma$ is constructed from convolutions of $\mG$ and bare interaction vertices. 

	\paragraph*{Dyson equation in real space.}
	Using this property, the first Dyson equation, Eq.~\eqref{eq:Dyson1}, becomes
	\begin{align}
		&\int \ddiff 2 \eu^{\iu \ch \left(\int_{r_2}^{r_1} + \int^{r_2}_{r_{1'}} \right) A(r')\cdot \diff r'} \left( \mG_0^{-1}\,' - \mSigma'\right)_{12}  \mG'_{2 1'} \non\\
		= \,& \eu^{\iu \ch \int_{r_{1'}}^{r_{1}} A(r')\cdot \diff r'} \delta_{1 1'}.
		\label{eq:DysonII1}
	\end{align}
	Moving the Wilson line on the RHS to the LHS, the product of Wilson lines depends only on the magnetic flux enclosed by the closed path $\mathcal{C}=(r_1 \rightarrow r_{1'} \rightarrow r_2 \rightarrow r_1)$: 
	\begin{align}
		\int \ddiff 2 \eu^{-\iu \frac{\theta_{ab}}{2}  (r_1 - r_2)_a(r_2 - r_{1'})_b} \left(\mG_0^{-1}\,' - \mSigma'\right)_{12}  \mG'_{2 1'} & =  \delta_{1 1'} ,
		\label{eq:DysonII2}
	\end{align}
	where we used
	\begin{align}
		\ch  \oint_{\mathcal{C}} A(r')\cdot \diff r' &= \ch B \frac{(r_2 - r_{1'})\times (r_1 - r_2)}{2} \non\\
		&=\frac{\theta_{ab}}{2} (r_2 - r_{1'})_a (r_1 - r_2)_b.
	\end{align}

	\paragraph*{Momentum-space formulation.}
	Next we perform a Fourier transformation for the bilocal functions, which depend only on the relative space-time coordinate. This replaces $(r_1 - r_2)$ in the exponential with $\iu \partial_p$ acting on $\left( \mG_0^{-1}\,' - \mSigma'\right)$ and $(r_2 - r_{1'})$ with $\iu \partial_p$ acting on $\mG'_{21'}$. Equation~\eqref{eq:DysonII2} then becomes
	\begin{align}
		\left( \mG_0^{-1}\,' - \mSigma'\right)_p \eu^{\iu \frac{\theta_{ab}}{2}  \overleftarrow{\partial}_{p_a} \overrightarrow{\partial}_{p_b}}  \mG'_p = 1,
		\label{eq:Dyson_iii}
	\end{align}
	which is precisely the Moyal-product form in Eq.~\eqref{eq:DysonFinal}. A detailed derivation using the Wigner transform~\cite{KamenevBook} is provided in App.~\ref{app:Dyson}. Physically, the Moyal algebra implies that the $a,b$ components of $p$ form a set of phase-space coordinates with symplectic two-form $\theta_{ab}$. First quantizing $p_a,p_b$ gives $[\hat{p}_a, \hat{p}_b]=\iu \theta_{ab}$, so $p_a,p_b$ are analogous to mechanical momentum. A more precise connection between $p$ and the first-quantized mechanical momentum appears in Sec.~\ref{sec:Omega1}.

	\paragraph*{Remarks.}
	The kernel $\mG_0^{-1}\,'$ depends only on the noninteracting Hamiltonian and is independent of $B$. Because of the Moyal algebra, even without interactions $(\mG'_p)^{-1} \neq \mG_0^{-1}\,'_p$. Expanding the Moyal algebra, $\mG'_p$ can be solved in powers of $B$. The expansion is perturbative only when $\omega_c/T \ll 1$. The self-energy $\mSigma'_p$ depends on $B$ implicitly and can in principle be computed self-consistently in powers of $B$ once a diagrammatic approximation is chosen together with Eq.~\eqref{eq:DysonII1}. For the purposes of this paper we show that the zero-field self-energy suffices to determine the orbital magnetization (for the class of approximations specified in Sec.~\ref{sec:Omega2}); finite-field corrections to $\mSigma'$ are left for future work.

	\section{Thermodynamic Potential}
	\label{sec:Omega}
	In this section we develop a systematic framework to compute the magnetic-field variation of the thermodynamic potential,
	$\frac{\delta \Omega_{\rm int}}{\delta B}$,
	and show it can be expressed entirely in terms of EM-gauge-invariant quantities: the Green's function $\mG'(p,\omega_n)$, the self-energy $\mSigma'(p,\omega_n)$, and the noninteracting kernel $\mG_0^{-1}\,'(p,\omega_n)$. This formulation enables computation of orbital magnetic response functions by solving the Dyson equations order by order in $B$.

	The thermodynamic potential is given by the Luttinger--Ward functional:
	\[
	\Omega_{\rm int} = -T \Tr \ln \mG^{-1} - T \Tr (\mSigma \circ \mG) + \Phi(\mG).
	\]
	Evaluating its magnetic-field dependence is challenging because of the Wilson-line structure and the logarithm. A central technical advance here is an operator-based expansion that resolves these difficulties.

	In Sec.~\ref{sec:Omega1} we reformulate the contribution from $-T \Tr \ln \mG^{-1}$ in terms of $\mathsf{H}_{p,n} = \mathsf{h}_{p}+\mSigma'(p,\omega_n)$. We find the $B$ field introduces gradient terms with open derivatives, which require careful handling. To proceed we introduce an operator representation of the thermodynamic potential to enable a systematic expansion in powers of $B$.
	In Sec.~\ref{sec:Omega2} we analyze the remaining terms, $-T \Tr (\mSigma \circ \mG) + \Phi(\mG)$.
	Additional remarks appear in Sec.~\ref{sec:OmegaRemarks}.
	\subsection{Contribution from $-T \Tr \ln \mG^{-1}$} 
	\label{sec:Omega1}

	From Eq.~\eqref{eq:Dyson1} we note that
	\[
	-T \Tr \ln \mG^{-1} = - T \Tr \ln (\mG^{-1}_0 - \mSigma).
	\]

	For conciseness, define
	\[
	\Omega^{(a)}_{\rm int} = -T \Tr \ln \mG^{-1}
	\]
	hereafter. Because of the logarithmic structure and the Wilson line, it is not immediate how to express this term in terms of $\mG_0^{-1}\,'$ and $\mSigma'$ in momentum space. To evaluate it we rewrite the term using the fermionic path integral:
	\begin{align}
		\Omega^{(a)}_{\rm int} =
		-T \ln \left(\int \Diff \bar{\psi} \Diff \psi \,
		\eu^{-\int \ddiff{1} \ddiff{2} \bar{\psi}_1 (\mG^{-1})_{12} \psi_{2}} \right),
		\label{eq:OmegaA0}
	\end{align}
	where $(\mG^{-1})_{12} = \eu^{\iu \ch \int_{r_2}^{r_1} A(r')\cdot \diff r'} \left(\mG_0^{-1}\,' - \mSigma'\right)_{12}$.

	To express the path integral in terms of $\mG_0^{-1}\,'$ and $\mSigma'$ in momentum space we need to absorb the Wilson line by transforming the fermion fields $\bar{\psi}$ and $\psi$. A naive local transformation on $\bar{\psi}_1$ and $\psi_{1'}$ is impractical. The key observation is that the Fourier label $p$ in Eq.~\eqref{eq:DysonFinal} corresponds physically to the mechanical momentum, as we noted in the discussion of Eq.~\eqref{eq:Dyson_iii}. This motivates defining a ``Fourier transformation'' based on noncommuting operators $\hat{p}$.

	\paragraph*{Generalized Fourier transformation.}
	We introduce an invertible transformation:
	\begin{align}
		{O}_{\hat{p}} = \int \diff^2 r\, \eu^{-\iu \hat{p} r} {O}_r, \quad
		{O}_r = \cTr \left(\eu^{\iu \hat{p} r}{O}_{\hat{p}}\right),
		\label{eq:generalizedF}
	\end{align}
	where $\cTr(\cdot)=\int \frac{\diff^2 \xi}{(2\pi)^2} \eu^{-\abs{\xi}^2/(2\theta)} \cbra{\xi} \cdot \cket{\xi}$ with coherent states $\cket{\xi}$ formed by $\hat{b}^{\dagger} = (\hat{p}_x - \iu \hat{p}_y)/\sqrt{2\theta}$. This basis accounts for the noncommutativity of $\hat{p}$. Following algebraic manipulations (see Appendix~\ref{appsec:FourierB}), the action becomes
	\begin{align}
		\Omega_{\rm int}^{(a)} = -T \ln \left(\int \Diff \bar{\psi} \Diff \psi \,
		\eu^{-\cTr\left(\bar{\psi}_{\hat{p}} (-\iu \omega_n - \mu + \mathsf{h}_{\hat{p}} + \mSigma'(\hat{p}, \omega_n)) \psi_{\hat{p}} \right)}\right),
		\label{eq:OmegaA1}
	\end{align}
	where $\mathsf{h}_{\hat{p}} = \sum_\delta (-\mathsf{t}_{\delta}) \eu^{- \iu \hat{p} \cdot \delta}$ and $\mSigma'(\hat{p}, \omega_n) = \int \diff \rho \,\eu^{-\iu \hat{p}\cdot \rho} \mSigma'(\rho, \omega_n)$.

	Next we express these fields in terms of commuting variables via the standard Wigner--Weyl transform. $\mathsf{h}_{\hat{p}}$ and $\mSigma'(\hat{p}, \omega_n)$ can be understood as the Weyl symbols of $\mathsf{h}_p$ and $\mSigma'(p, \omega_n)$~\cite{Szabo2003,Douglas2001rmp}, which are
	\begin{align}
		\mathsf{h}_{{p}} &= \sum_\delta (-\mathsf{t}_{\delta}) \eu^{- \iu p \cdot \delta},\non\\
		\mSigma'(p, \omega_n) &= \int \diff \rho \,\eu^{-\iu {p}\cdot \rho} \mSigma'(\rho, \omega_n).
	\end{align}
	The path integral can be written as
	\begin{align}
		\Omega_{\rm int}^{(a)} =
		-T \ln \left( \int \Diff \bar{\psi}_p \Diff \psi_p \,
		\eu^{- \sum_n \int \frac{\diff^2 p}{(2\pi)^2} \bar{\psi}_p \star (-\iu \omega_n -\mu + \mathsf{H}_{p,n}) \star \psi_p}\right),
		\label{eq:OmegaA2}
	\end{align}
	where $\mathsf{H}_{p,n} = \mathsf{h}_{p} + \mSigma'(p,\omega_n)$, and $\star$ is the Moyal algebra defined previously. It can be checked that the Wigner--Weyl transform does not introduce a nontrivial Jacobian into the path integral measure. 

	Integrating out the fermion fields yields
	\begin{widetext}
		\begin{align}
			\Omega_{\rm int}^{(a)} &=
			-T \Tr \ln \Big[\beta \big(-\iu \omega_n - \mu + \mathsf{H}_{p,n} \eu^{\frac{\iu \theta_{ab}}{2}\overleftarrow{\partial_a}\overrightarrow{\partial_b}} \big)(2\pi)^2\delta(p-p')\Big]\non\\
			&= -T \Tr \ln \Big[\beta \big(-\iu \omega_n - \mu + \mathsf{H}_{p,n}+\sum_{m=1}^{\infty} \frac{1}{m!} \big(\prod_{i=1}^m \frac{\iu \theta_{a_i b_i}}{2}\big) (\partial_{a_1}\cdots\partial_{a_m}\mathsf{H}_{p,n}) \partial_{b_1}\cdots\partial_{b_m} \big)(2\pi)^2\delta(p-p')\Big].
			\label{eq:OmegaA22}
		\end{align}
	\end{widetext}
	Here $\partial_{a,b} = \partial_{p_a,p_b}$, and open derivatives act on $\delta(p-p')$. Expanding in powers of $B$ requires accounting for both gradient terms and $\mSigma'$. While $\mSigma'$ can be obtained from self-consistent Dyson solutions, the gradient expansion is subtle because of the open derivatives and the logarithm.

	\paragraph*{Operator formulation and logarithmic expansion.}
	To proceed, it is convenient to define operators $\widehat{\mathsf{H}}_n$ and $\widehat{\mathsf{X}}$ acting on the momentum-space basis for single-particle wave functions, $\{\psi_p\}$:
	\begin{align}
		\widehat{\mathsf{H}}_n \ket{\psi_{p}} &=  \ket{\psi_{p}} \mathsf{H}_{p,n}, \non\\
		\widehat{\mathsf{X}} \ket{\psi_{p}} &= \int \frac{\diff^2 p'}{(2\pi)^2} \ket{\psi_{p'}} \iu \partial_{p'} (2\pi)^2\delta(p-p').
		\label{eq:HXv}
	\end{align}
	Then
	\begin{widetext}
		\begin{align}
			\Omega_{\rm int}^{(a)}
			= -T \Tr \ln \beta \Big(-\iu \omega_n - \mu + \widehat{\mathsf{H}}_n+\sum_{m=1}^{\infty} \frac{1}{m!} \big(\prod_{i=1}^m \frac{-\iu \theta_{a_i b_i}}{2}\big) [\widehat{\mathsf{X}}_{a_m},\dots,[\widehat{\mathsf{X}}_{a_1}, \widehat{\mathsf{H}}_{n}]] \widehat{\mathsf{X}}_{b_1}\cdots\widehat{\mathsf{X}}_{b_m} \Big).
			\label{eq:OmegaA3}
		\end{align}
	\end{widetext}
	To expand the logarithm in powers of $B\propto \theta$, note the zeroth-order term $(-\iu \omega_n - \mu + \widehat{\mathsf{H}}_n)$ does not commute with the remainder. To expand the logarithm for noncommuting operators we use 
	\begin{align}
		& \ln (\widehat{A}+\theta\widehat{B}) =\non\\
		& \quad \ln \widehat{A} + \int_0^\infty \diff z (\widehat{A}+z \mathbb{I})^{-1} (\theta \widehat{B}) (\widehat{A}+z \mathbb{I})^{-1} \non\\
		& \quad-  \int_0^\infty \diff z (\widehat{A}+z \mathbb{I})^{-1} (\theta \widehat{B}) (\widehat{A}+z \mathbb{I})^{-1} (\theta \widehat{B}) (\widehat{A}+z \mathbb{I})^{-1} \non\\
		& \quad+ \mathcal{O}(\theta^3),
		\label{eq:LogExpansion}
	\end{align}
	which follows from $\int_0^x \diff z (\widehat{A}+\theta\widehat{B}+z)^{-1} = \ln (\widehat{A}+\theta \widehat{B} +z)|_0^{x}$ and the Taylor expansion of $(\widehat{A}+ \theta \widehat{B})^{-1}$. We emphasize that because the perturbation $\theta \widehat{B}$ contains open derivatives in this case, the cyclic trace property does not apply in the naive way. More detailed discussion of this subtlety appears in Sec.~\ref{sec:Omega3}, where we apply the formalism to the $B$-linear term of $\Omega_{\rm int}$.

	\subsection{Contribution from $-T \Tr \mSigma \circ \mG + \Phi(\mG)$}
	\label{sec:Omega2}

	For conciseness define
	\[
	\Omega^{(b)}_{\rm int} = -T \Tr (\mSigma \circ \mG) + \Phi(\mG).
	\]
	As for $\Omega^{(a)}_{\rm int}$ in Sec.~\ref{sec:Omega1}, we first express this term through the EM-gauge-invariant functions $\mG'$ and $\mSigma'$ and then differentiate with respect to $B$. The magnetic field then enters in two ways: implicitly, through $\mG'$ and $\mSigma'$, and explicitly, through the flux factors that the Wilson lines leave behind.

	\paragraph*{Trace term.} Using Eq.~\eqref{eq:bilocal} for $\mSigma$ and $\mG$,
	\begin{align}
		\Tr(\mSigma\circ\mG) = \int \ddiff{1}\ddiff{2}\, \eu^{\iu\ch\left(\int_{r_2}^{r_1}+\int_{r_1}^{r_2}\right)A(r')\cdot\diff r'}\, \tr\!\left[\mSigma'_{12}\,\mG'_{21}\right].
	\end{align}
	The two Wilson lines run along the same straight segment in opposite directions and cancel. Fourier transforming the remaining translationally invariant integrand gives
	\begin{align}
		-T\Tr(\mSigma\circ\mG) = -T\mathcal{V}\sum_n\int\frac{\diff^2p}{(2\pi)^2}\,\tr\!\left[\mSigma'(p,\omega_n)\,\mG'(p,\omega_n)\right].
		\label{eq:SigmaGprime}
	\end{align}
	This term has no explicit $B$ dependence.

	\paragraph*{The 2PI functional $\Phi$.} Each skeleton diagram of $\Phi(\mG)$ consists of closed fermion loops connected by interaction lines. We consider density--density interactions between orbitals, whose bare vertices carry no Wilson lines. For a fermion loop $L$ with vertices $1,\dots,k$, Eq.~\eqref{eq:bilocal} gives
	\begin{align}
		\mG_{12}\,\mG_{23}\cdots\mG_{k1} = \eu^{\iu\ch\oint_{\mathcal{C}_L}A(r')\cdot\diff r'}\;\mG'_{12}\,\mG'_{23}\cdots\mG'_{k1},
	\end{align}
	where $\mathcal{C}_L$ is the closed polygon $r_1\to r_k\to\cdots\to r_2\to r_1$ formed by the straight Wilson lines. By Stokes' theorem,
	\begin{align}
		\ch\oint_{\mathcal{C}_L}A(r')\cdot\diff r' = \ch B\,\mathcal{A}_L,
		\label{eq:loopflux}
	\end{align}
	with $\mathcal{A}_L$ the oriented area enclosed by $\mathcal{C}_L$. The flux factor is EM-gauge invariant: it is the Aharonov--Bohm phase of the loop. Hence
	\begin{align}
		\Phi(\mG) = \Phi'(\mG';B),
		\label{eq:PhiPrime}
	\end{align}
	where $\Phi'$ is given by the same skeleton diagrams evaluated with $\mG'$ and with a factor $\eu^{\iu\ch B\mathcal{A}_L}$ for every fermion loop. Thus $\Phi'$ depends on $B$ implicitly through $\mG'$ and explicitly through the flux factors. A loop whose vertices occupy at most two distinct positions encloses no area. This holds in self-consistent Hartree--Fock, in self-consistent RPA ($\Phi$ given by the ring series of polarization bubbles built from the dressed $\mG$, known as the GW approximation in the electronic-structure literature), and in single-site DMFT; for these approximations $\Phi'$ has no explicit $B$ dependence. We will consider this \emph{no-flux} approximation hereafter.

	\paragraph*{Field derivative.} By Eqs.~\eqref{eq:SigmaGprime} and~\eqref{eq:PhiPrime}, $\Omega^{(b)}_{\rm int}$ depends on $B$ through $\mSigma'$, $\mG'$, and the flux factors. Differentiating,
	\begin{align}
		\frac{\delta (-T \Tr \mSigma \circ \mG)}{\delta B} = {} &-T\mathcal{V}\sum_n\int\frac{\diff^2p}{(2\pi)^2}\,\tr\!\left[\frac{\delta\mSigma'}{\delta B}\,\mG'\right] \non\\
		&{}-T\mathcal{V}\sum_n\int\frac{\diff^2p}{(2\pi)^2}\,\tr\!\left[\mSigma'\,\frac{\delta\mG'}{\delta B}\right], \non\\
		\frac{\delta \Phi'(\mG';B)}{\delta B} = {} &T\mathcal{V}\sum_n\int\frac{\diff^2p}{(2\pi)^2}\,\tr\!\left[\mSigma'\,\frac{\delta\mG'}{\delta B}\right] + \partial_B\Phi'\big|_{\mG'}.
	\end{align}
	The first term of $\delta\Phi'/\delta B$ is the variation through $\mG'$. The second term, $\partial_B\Phi'|_{\mG'}$, is the derivative with respect to the explicit $B$ in the flux factors at fixed $\mG'$, and it vanishes in the no-flux approximation. The terms with $\delta\mG'/\delta B$ cancel, and the total contribution simplifies to
	\begin{align}
		\frac{\delta \Omega^{(b)}_{\rm int}}{\delta B} = -T\mathcal{V}\sum_n\int\frac{\diff^2p}{(2\pi)^2}\,\tr\!\left[\frac{\delta\mSigma'(p,\omega_n)}{\delta B}\,\mG'(p,\omega_n)\right].
		\label{eq:OmegaB}
	\end{align}

	\subsection{Additional remarks}
	\label{sec:OmegaRemarks}

	We summarize several key points following from Eqs.~\eqref{eq:OmegaA3} and~\eqref{eq:OmegaB}.

	First, the thermodynamic potential $\Omega_{\rm int}$ depends on the magnetic field $B$ in two distinct ways. There is an explicit dependence arising from the operator expansion in the noncommutativity parameter $\abs{\theta_{ab}}\propto B$ in Eq.~\eqref{eq:OmegaA3}, as well as an implicit dependence through the Green's function $\mG'$ and self-energy $\mSigma'$, which are determined self-consistently by the Dyson equations reviewed in Sec.~\ref{sec:DysonMorb}. In general, $\Phi'$ carries an additional explicit dependence on $B$ (see Sec.~\ref{sec:Omega2}).

	Second, within this formulation the magnetic field enters exclusively through gradient operators acting in the effective momentum space, as dictated by the Moyal algebra. In particular, the field does not induce any explicit time dependence or drive the system out of equilibrium. This is consistent with the physical distinction between magnetic and electric fields: a static magnetic field modifies the phase-space structure of electronic states while preserving thermodynamic equilibrium.

	Finally, for the logarithmic expansion in Eq.~\eqref{eq:LogExpansion} to be controlled, the perturbative parameter must satisfy $\abs{\theta \widehat{B}}/\abs{\widehat{A}} \ll 1$. Applied to the present problem, this condition translates into $\omega_c/T \ll 1$, where $\omega_c$ is the cyclotron frequency, which depends on both the effective mass and quantum-geometric quantities for multiband systems~\cite{Ethan-Ye-Wang}. In this regime, oscillatory contributions associated with Landau quantization are exponentially suppressed and do not appear in the perturbative expansion of the thermodynamic potential, allowing smooth orbital magnetic response coefficients to be defined order by order in $B$.

	\section{Application to Spontaneous Orbital Magnetization}
	\label{sec:Omega3}

	We apply the framework above to compute the spontaneous orbital magnetization, which corresponds to the $B$-linear coefficient of $\Omega_{\rm int}$, or equivalently the zeroth-order term of $\frac{\delta \Omega_{\rm int}}{\delta B}$ [see Eq.~\eqref{eq:Morb_def}].

	We first outline the derivation in Sec.~\ref{sec:Morb_derivation} and show $M_{\rm orb}$ can be fully determined by the renormalized zero-field Hamiltonian $\mathsf{H}^{(0)}_{p,n} = \mathsf{h}_{p} + \mSigma'^{(0)}(p, \omega_n)$. In Sec.~\ref{sec:Morb_discussion} we treat the case when $\mathsf{H}^{(0)}_{p,n}$ is Hermitian and discuss the physical interpretation and connections to earlier mean-field results.

	\subsection{Derivation of $M_{\rm orb}$}
	\label{sec:Morb_derivation}
	The solutions to the Dyson equations can be expanded in powers of $B$:
	\begin{align}
		\mSigma'(p,\omega_n) &= \sum_{l=0}^\infty B^l\, \mSigma'^{(l)}(p,\omega_n), \non\\
		\mG'(p,\omega_n) &= \sum_{l=0}^\infty B^l\, \mG'^{(l)}(p,\omega_n),
	\end{align}
	where $\mSigma'^{(0)}(p,\omega_n)$ and $\mG'^{(0)}(p,\omega_n)$ are solutions to the Dyson equations at $B=0$.
	The effective Hamiltonian $\mathsf{H}_{p,n}$ also follows from this expansion as $\mathsf{H}^{(0)}_{p,n} = \mathsf{h}_p + \mSigma'^{(0)} (p,\omega_n)$ and $\mathsf{H}^{(l\geqslant 1)}_{p,n} = \mSigma'^{(l)}(p,\omega_n)$. To find $B$-linear coefficients in $\Omega_{\rm int}$ it suffices to keep terms in $\mSigma'^{(l)},\, \mG'^{(l)}$ up to $l=1$.

	Using Eqs.~\eqref{eq:OmegaA3} and \eqref{eq:LogExpansion}, and absorbing the chemical potential $\mu$ into $\widehat{\mathsf{H}}_n$, the contribution from $-T \Tr \ln \mG^{-1}$ becomes:
	\begin{widetext}
		\begin{align}
			\Omega^{(a)}_{\rm int} &= -T \Tr \ln \beta\big(-\iu \omega_n + \widehat{\mathsf{H}}^{(0)}_n + B\widehat{\mSigma}'^{(1)}_n + \tfrac{\ch B \epsilon_{ab}}{2} \widehat{\mv}^{(0)}_a \widehat{\mX}_b\big) + \dots \non\\
			&= -T \Tr \ln \beta(-\iu \omega_n + \widehat{\mathsf{H}}^{(0)}_n) \non\\
			&\quad -B T \Tr \int_0^\infty \diff z \frac{1}{\beta(-\iu \omega_n + \widehat{\mathsf{H}}^{(0)}_n)+z} \beta\Big(\widehat{\mSigma}'^{(1)}_n + \tfrac{\ch \epsilon_{ab}}{2}\widehat{\mv}^{(0)}_a \widehat{\mX}_b\Big) \frac{1}{\beta(-\iu \omega_n + \widehat{\mathsf{H}}^{(0)}_n)+z}.
		\end{align}
	\end{widetext}
	Here $\widehat{\mv}^{(0)}_a$ is defined like $\widehat{\mathsf{H}}_n$ in Eq.~\eqref{eq:HXv}, with $\mathsf{H}_{p,n}$ replaced by the generalized velocity $\mv^{(0)}_a=\partial_{p_a}\mathsf{H}^{(0)}_{p,n}$.

	The term involving $\widehat{\mSigma}'^{(1)}_n$ simplifies via cyclic trace. Integrating over $z$, we find
	\begin{align}
		\Omega_{\rm int}^{(a,1)} &= BT \Tr\!\left[\frac{1}{\iu \omega_n - \widehat{\mathsf{H}}^{(0)}_n}\widehat{\mSigma}'^{(1)}_n\right] \non\\
		&= BT \mathcal{V} \sum_n \int \frac{\diff^2 p}{(2\pi)^2} \tr\!\big[\mSigma'^{(1)}(p,\omega_n)\mG'^{(0)}(p,\omega_n)\big].
	\end{align}

	The term involving $\widehat{\mv}^{(0)}_a \widehat{\mX}_b$ requires care since $\widehat{\mX}$ acts as an open derivative. Inserting $\mathbb{1} = \int_p \ket{\psi_p}\bra{\psi_p}$ formed by a complete basis in orbital space and using Eq.~\eqref{eq:HXv}, the $B$-linear contribution becomes:
	\begin{widetext}
		\begin{align}
			\Omega_{\rm int}^{(a,2)} &= -\iu \frac{\theta_{ab}}{2}\mathcal{V}\sum_n \int \frac{\diff^2 p}{(2\pi)^2}\int_0^\infty \diff z\, \tr\!\Bigg[\frac{1}{\beta(-\iu \omega_n+\mathsf{H}^{(0)}_{p,n})+z}(\partial_a\mathsf{H}^{(0)}_{p,n})\partial_b\Bigg(\frac{1}{\beta(-\iu \omega_n+\mathsf{H}^{(0)}_{p,n})+z}\Bigg)\Bigg].
		\end{align}
	\end{widetext}
	Here the volume $\mathcal{V}$ arises from $\int_{p,p'} (\delta(p-p'))^2$.

	From Eq.~\eqref{eq:OmegaB}, the contribution from $\Omega_{\rm int}^{(b)}$ is
	\[
	\Omega_{\rm int}^{(b)} = -BT\mathcal{V}\sum_n\int\frac{\diff^2 p}{(2\pi)^2}\tr\!\big[\mSigma'^{(1)}(p,\omega_n)\mG'^{(0)}(p,\omega_n)\big],
	\]
	which cancels $\Omega_{\rm int}^{(a,1)}$. Thus $M_{\rm orb}\propto\Omega_{\rm int}^{(a,2)}$, and
	\begin{widetext}
		\begin{align}
			M_{\rm orb} &= -\frac{\iu e}{2}\epsilon_{ab}\sum_n\int\frac{\diff^2 p}{(2\pi)^2}\int_0^\infty\diff z\,\tr\!\Bigg[\frac{1}{\beta(-\iu \omega_n+\mathsf{H}^{(0)}_{p,n})+z}(\partial_a\mathsf{H}^{(0)}_{p,n})\partial_b\Bigg(\frac{1}{\beta(-\iu \omega_n+\mathsf{H}^{(0)}_{p,n})+z}\Bigg)\Bigg].
			\label{eq:Morb_v}
		\end{align}
	\end{widetext}
	Here we set $\ch=-e$ for electrons. Equation~\eqref{eq:Morb_v} applies to interacting systems with frequency-dependent self-energies in the no-flux approximation. It shows the magnetic field couples to the noninteracting Hamiltonian and self-energy in the same structural way. 

	\subsection{Discussion}
	\label{sec:Morb_discussion}
	It is useful to express the result in terms of eigenstates and eigenenergies of the renormalized Hamiltonian. We consider cases where $\mathsf{H}^{(0)}_{p,n}$ is Hermitian, so $\mathsf{H}^{(0)}_{p,n} = \mathsf{V}_{p,n} \mathsf{\Lambda}_{p,n} \mathsf{V}^\dagger_{p,n}$, with $\mathsf{V}_{p,n}$ unitary and $\mathsf{\Lambda}_{p,n}$ real and diagonal. For brevity we drop the subscript $p,n$ unless needed. We find that $\partial \mathsf{\Lambda}$ contributions to $M_{\rm orb}$ vanish, leaving only geometrical terms of the eigenstates:
	\begin{widetext}
		\begin{align}
			M_{\rm orb} & =  - \iu \frac{e }{2}\epsilon_{ab} \sum_n \int \frac{\diff^2 p}{(2\pi)^2} \int_0^{\infty}\diff z  \left\{\tr\left[\left(\frac{1}{\beta(-\iu \omega_n + {\mathsf{\Lambda}})+z}\right)^2 (\mV^{\dagger} \partial_a \mV \mathsf{\Lambda} - \mathsf{\Lambda} \mV^{\dagger} \partial_a \mV) \mV^\dagger  \partial_b \mV \right] + \right. \non\\
			&  \qquad \qquad\qquad\qquad\qquad\qquad\quad \left. \tr\left[ \frac{1}{\beta(-\iu \omega_n + {\mathsf{\Lambda}})+z} (\mV^{\dagger} \partial_a \mV \mathsf{\Lambda} - \mathsf{\Lambda} \mV^{\dagger} \partial_a \mV) \frac{1}{\beta(-\iu \omega_n + {\mathsf{\Lambda}})+z} \partial_b \mV^{\dagger} \mV\right] \right\}
		\end{align}
	\end{widetext}
	Each of the two bracketed terms is invariant under the gauge transformation 
	\begin{align}
		\mV \rightarrow \mV \mathsf{\Theta},
		\label{eq:gauge}
	\end{align}
	with $\mathsf{\Theta}$ diagonal and unitary. These two terms correspond to distinct physical processes.

	\paragraph*{Magnetic moment contribution.}
	Evaluating the $z$ integral in the first term yields the Green's function expressed in the band basis. Using the identity	$\mathsf{v}_{a}^{(0)} = \mathsf{V}\left(\mV^{\dagger} \partial_a \mV \mathsf{\Lambda} - \mathsf{\Lambda} \mV^{\dagger} \partial_a \mV\right)\mathsf{V}^{\dagger} + \mathsf{V}(\partial_a\mathsf{\Lambda})\mathsf{V}^{\dagger}$ (the last term does not contribute),
	together with $\widehat{\mX}_b \sim \partial_b$, this contribution can be interpreted diagrammatically as a tadpole term with a vertex determined by a band-projected, momentum-frequency-resolved orbital magnetic dipole moment. Explicitly, we define
	\begin{align}
		\mathfrak{m}^{\lband}(p, \omega_n) & = \frac{\iu }{2}\bra{u^{\lband}} \epsilon_{ab}\widehat{\mathsf{v}}_{a}^{(0)} \widehat{\mX}_{b} \ket{u^{\lband}}\non\\
		& = \frac{\iu }{2} \epsilon_{ab} 
		\ip{\partial_a u^{\lband}}{u^{\mband}}
		\left( \lambda^{\lband} - \lambda^{\mband} \right)
		\ip{u^{\mband}}{\partial_b u^{\lband}}.
		\label{eq:orbmoment}
	\end{align}
	Here $\lband$ and $\mband$ label band indices, and repeated band indices $\mband$ are implicitly summed. The frequency-dependent Bloch wave functions are defined through
	\begin{align}
		\left( \ket{u^1_{p,n}}, \ket{u^2_{p,n}}, \ldots, \ket{u^N_{p,n}} \right) = \mV_{p,n}.
	\end{align}

	\paragraph*{Berry curvature contribution.}
	The second term is nonzero only for distinct bands in the resolvents. After integrating out $z$, the $\mathsf{\Lambda}$ in the numerator cancels with resolvents, leaving a truly geometrical result determined by the momentum-frequency-resolved Abelian Berry curvature 
	\begin{align}
	\mathfrak{b}^{\lband} (p,\omega_n) = \iu \epsilon_{ab} \ip{\partial_a u^{\lband}}{\partial_b u^{\lband}}
	\label{eq:berrycurvature}
	\end{align}
	and the eigenenergy of the band. 

	Specifically, in terms of the momentum-frequency-dependent eigenenergies $\lambda^{\lband}(p,\omega_n)$, orbital moments $\mathfrak{m}^{\lband}(p,\omega_n)$, and Berry curvatures $\mathfrak{b}^{\lband}(p,\omega_n)$, $M_{\rm orb}$ reads
	\begin{align}
		&M_{\rm orb}
		=  e T \sum_n \int \frac{\diff^2 p}{(2\pi)^2} \sum_{\lband}  \non\\
		&   \quad \left(\frac{1}{\iu \omega_n - \lambda^{\lband}}  \mathfrak{m}^{\lband} (p,\omega_n) +     \left( \ln \beta(-\iu \omega_n +\lambda^{\lband}) \right)  \mathfrak{b}^{\lband} (p,\omega_n)\right).
		\label{eq:Morb_hermitian}
	\end{align}
	For frequency-independent self-energies the Matsubara sum can be performed as in the noninteracting case, and we recover the noninteracting formula with a renormalized Hamiltonian:
	\begin{widetext}
		\begin{align}
			M_{\rm orb} = \iu e \epsilon_{ab}\sum_{\lband}\int\frac{\diff^2 p}{(2\pi)^2}\Big[\frac{1}{2}\bra{\partial_a u^{\lband}_p}(\lambda^{\lband}_p-\mathsf{H}_p)\ket{\partial_b u^{\lband}_p}f_F(\lambda^{\lband}_p)+T\ip{\partial_a u^{\lband}_p}{\partial_b u^{\lband}_p}\ln\big(1+\eu^{-\beta\lambda^{\lband}_p}\big)\Big].
		\end{align}
	\end{widetext}

\section{Summary and Outlook}
\label{sec:summary_outlook}

\subsection{Summary}
\label{sec:summary}

In this work we developed a quantum many-body framework for orbital magnetic responses in interacting multiband electron systems. Starting from the Luttinger--Ward functional, we formulated the thermodynamic potential in the presence of a weak, static magnetic field and constructed a controlled expansion in powers of $B$ in the regime $\omega_c/T \ll 1$, where oscillatory Landau level effects are exponentially suppressed. A central technical ingredient is a modified ``Fourier'' representation based on noncommuting mechanical-momentum operators, which maps the problem to an effective momentum space governed by the Moyal algebra. Within this representation, the magnetic field enters locally through gradient operators, enabling a systematic treatment of the trace-log term in the thermodynamic potential.

Using this formulation, we showed that the magnetic-field dependence of the thermodynamic potential arises from two sources: an explicit dependence encoded in the Moyal expansion and an implicit dependence through the Green's function and self-energy determined self-consistently by the Dyson equations. At linear order in $B$, the latter contribution cancels as a consequence of the stationarity of the Luttinger--Ward functional, leaving a compact expression controlled entirely by zero-field quantities. As a result, the spontaneous orbital magnetization $M_{\rm orb}$ can be expressed solely in terms of the renormalized zero-field Hamiltonian, given by the sum of the bare Bloch Hamiltonian and the zero-field self-energy. This holds when the 2PI functional, written in terms of $\mG'$, carries no explicit $B$ dependence, as for Hartree--Fock, self-consistent RPA, and single-site DMFT; the general case, in which fermion loops enclosing magnetic flux contribute such a dependence, will be discussed elsewhere.

For Hermitian renormalized Hamiltonians, we further decomposed $M_{\rm orb}$ into two gauge-invariant contributions defined in momentum-frequency space: a band-resolved orbital magnetic moment and a Berry-curvature type term. This structure generalizes the modern theory of orbital magnetization to interacting systems with frequency-dependent self-energies. In the static self-energy limit, our results reduce smoothly to the well-known geometric formula for noninteracting electrons, with the noninteracting Hamiltonian replaced by a self-energy--renormalized Hamiltonian.

A key approximation underlying the present formulation is the weak-field continuum limit, in which the magnetic length is much larger than the lattice spacing, $\abs{\ch B}a^2 \ll 1$. While this is a standard assumption in the theory of orbital magnetic responses~\cite{Kohn1959,Blount1962,Roth1962,Roth1966}, it neglects the discreteness of lattice translations. In this regime the discrete magnetic translation symmetry is effectively enlarged to a continuous magnetic translation symmetry, and the Landau-level degeneracy is determined solely by the total magnetic flux through the system~\cite{ye_wang_first}. This approximation is distinct from the condition $\omega_c/T \ll 1$, which controls the perturbative expansion of the thermodynamic potential and the suppression of quantum oscillations. 

\subsection{Outlook}
\label{sec:outlook}

The formalism developed here provides a general platform for analyzing orbital magnetic responses of correlated single-band and multiband systems using self-consistent solutions of the Dyson equations, as obtained from diagrammatic or numerical many-body approaches. In particular, it can be directly applied to access higher-order orbital magnetic responses through a systematic expansion of the thermodynamic potential in powers of the magnetic field~\cite{ye_susceptibility}. As a benchmark, the same expansion reproduces the Landau--Peierls orbital susceptibility for a single noninteracting band with arbitrary dispersion.


Several extensions of the present framework constitute natural directions for future work. One is the application to superconducting states, where anomalous Green's functions and pairing fields must be incorporated in an EM-gauge-consistent formulation of the Luttinger--Ward functional. Another is the inclusion of disorder, which would require extending the present thermodynamic formulation to incorporate impurity scattering in a manner compatible with gauge covariance and magnetic translations. Finally, relaxing the continuum approximation and explicitly treating the discrete magnetic translation group may be necessary to access regimes of stronger magnetic fields or lattice-scale magnetic phenomena.

\section*{Acknowledgments}
I thank Jing-Yuan Chen, Andrey Chubukov, Zhihuan Dong, Duncan Haldane, Chao-Ming Jian, Jian Kang, Guangjie Li, Dan Mao, Sri Raghu, Oskar Vafek, Xiaoyu Wang, Xiaochuan Wu, and Yi-Ming Wu for helpful discussions related to this work. I am especially grateful to Leon Balents for his encouragement, support, and valuable feedback. I also thank Yuxuan Wang for collaborations on related problems and for many useful discussions.

The early stages of this work were carried out during a postdoctoral appointment at the Kavli Institute for Theoretical Physics, which is supported by the Simons Foundation (Grant No.~216179, LB) and by the National Science Foundation under Grant No.~NSF PHY-1748958. I also acknowledge the hospitality of the Aspen Center for Physics, which is supported by the National Science Foundation under Grant No.~PHY-1607611.

\bibliography{OrbMagBib}


\onecolumngrid
\appendix
\newpage
Here is an outline of the Appendices. We summarize the notations and conventions in App.~\ref{app:notation}. We derive the Dyson equation using the Weyl (Wigner--Weyl) transform in App.~\ref{app:Dyson}. A brief review of the Luttinger--Ward functional is given in App.~\ref{app:LWintroduction}, where we explicitly specify the ordering conventions for bilocal coordinates for clarity. In App.~\ref{app:Fourier}, we present technical details of the generalized Fourier transformation, which is crucial for expressing the fermionic path integral in an effective momentum representation.

\section{Notation}
\label{app:notation}
In this paper we consider a two-dimensional (2d) system with a magnetic field perpendicular to the plane and along the $z$ axis.

{\bf Notations for variables:}
\begin{itemize}
	\item Vectors in real or momentum space: $A, r, k, p,\tilde{p}, v,R,\bar{R},\nabla$.
	\item Indices for vectors in real or momentum space: $a, b, c, ...$.
	\item Basis of vector space: $\ve{x}, \ve{y}, \ve{z}$.
	\item Operators in first quantization: $\hat{O}$.
	\item Operators in second quantization: $\widehat{O}$.
	\item Matsubara-frequency indices: $\omega_n,\omega_{n'}, ...$.
	\item Matrix-valued functions in orbital or band space: $\mathsf{t}, \mathsf{H}, \mathsf{\Lambda},...$.
	\item Orbital indices: $\alpha, \beta, ...$.
	\item Band indices: $\mband, \lband, ...$.
\end{itemize}

{\bf Momentum and coordinates for a charged particle in a magnetic field:}
Consider a particle with charge $\ch$ ($\ch = -e$ for electrons) in a magnetic field ${\bf B}=B \ve{z}$. The gauge field, relevant momenta and coordinates for classical motion, and their quantizations are:
\begin{itemize}
	\item In the symmetric gauge, ${ A}=\frac{B}{2} (\ve{z} \times r)_\perp=\frac{B}{2} (-y,x)$.
	\item The canonical momentum and the dual momentum are
	\begin{equation}
		{k} = - \iu {\nabla},\qquad
		{p} = {k}- \ch {A} = - \iu {\nabla} - \ch {A}, \qquad
		{\tilde{p}} = {k} + \ch {A} = - \iu {\nabla} + \ch {A}.
	\end{equation}
	After \emph{first} quantization, the commutators read
	\begin{equation}
		[\hat{p}_a, \hat{p}_b] = \iu \theta_{ab},\qquad [\hat{\tilde{p}}_a, \hat{\tilde{p}}_b] =- \iu \theta_{ab},
	\end{equation}
	where $\theta_{ab}=\ell_B^{-2}\epsilon_{ab}$, with $\ell_B$ the magnetic length, defined as $\ell_B^{-2}=\ch B$.
	\item We also define the corresponding coordinates. The classical orbit coordinate (relative to the guiding center) is
	\begin{equation}
		\bar{{R}}=- \ell_B^2 ({p}\times \ve{z})_\perp, \qquad
		\bar{R}_a = - \ell_B^2 \epsilon_{ab} p_b
		\Longrightarrow
		[\hat{\bar{R}}_a,\hat{\bar{R}}_b]=\iu \epsilon_{ab} \ell_B^2.
	\end{equation}
	The guiding-center coordinate is defined as
	\begin{equation}
		R_a = r_a - \bar{R}_a
		\Longrightarrow
		{R}= \ell_B^2( \tilde{p} \times \ve{z})_\perp
		\Longrightarrow
		[\hat{R}_a, \hat{R}_b]= -\iu \epsilon_{ab} \ell_B^2.
	\end{equation}
	For $\ch>0$ the particle rotates clockwise, consistent with the classical orbit velocity ${ v}=\frac{\ch B}{m} ({ r}\times {\bf z})_\perp$.
\end{itemize}

{\bf Conventions for Fourier transformation:}
\begin{align}
	{\text{Discrete momentum}} &\Longleftrightarrow {\text{Continuous momentum}}\non\\
	\sum_k &\Longleftrightarrow \mathcal{V} \int \frac{\diff k}{(2\pi)^d},\non\\
	\delta_k &\Longleftrightarrow \frac{(2\pi)^d}{\mathcal{V}} \delta (k),\non\\
	\frac{1}{\mathcal{V}} \int \diff r\, \eu^{\iu k r} = \delta_k &\Longleftrightarrow \int \diff r\, \eu^{\iu k r} = (2\pi)^d \delta (k)\non\\
	a_r = \frac{1}{\sqrt{\mathcal{V}}}\sum_k \eu^{\iu k r}a_k & \Longleftrightarrow a_r =\sqrt{\mathcal{V}} \int \frac{\diff k}{(2\pi)^d} \eu^{\iu k r}a_k \text{ with } a_k = \frac{1}{\sqrt{\mathcal{V}}} \int \diff r\, \eu^{-\iu k r} a_r \non\\
	a_i = \frac{1}{\sqrt{\mathcal{N}}}\sum_k \eu^{\iu k r_i}a_k & \Longleftrightarrow a_i =\sqrt{\mathcal{N}} \int \frac{\diff k}{(2\pi)^d/v} \eu^{\iu k r_i}a_k \text{ with } a_k = \frac{1}{\sqrt{\mathcal{N}}} \sum_i\, \eu^{-\iu k r_i} a_i 
\end{align}
Here, $\mathcal{V}$ is the total volume, and $v$ is the volume of a unit cell. The operators $a_k$ with discrete and continuous $k$ have the same physical units. The real-space fields $a_r$ and lattice operators $a_i$ differ in normalization, with $a_i = \sqrt{v}\, a_r$.

\section{Dyson equation}
\label{app:Dyson}
In this section, we derive the first Dyson equation [Eq.~\eqref{eq:Dyson1}] in momentum-frequency representation. For definiteness we work in the symmetric gauge.

The first Dyson equation can be written as
\begin{align}
	\int \ddiff 2 \mG_0^{-1}(1,2)  \mG(2,1') = \int \ddiff 2 \mSigma(1,2) \mG(2,1') + \delta_{1 1'} ,
	\label{appeq:Dyson0}
\end{align}
where
$\mG_0^{-1} (1,2) = \delta(\tau_1-\tau_2)\delta(r_1-r_2) (-\partial_{\tau_2}+\mu) - \delta(\tau_1- \tau_2)  \sum_{\delta} \delta({r_1-\delta- r_2}) (- {\mathsf{t}_{\delta}}) \eu^{\iu \ch {A}_{(r_1 + r_2 )/2} \cdot (r_1-r_2)}$.

Define the relative and CoM coordinates for $1,1'$ as
\begin{align}
	T &= \frac{\tau_1 + \tau_{1'}}{2}, \, {R} = \frac{r_1 + r_{1'}}{2}\non\\
	t &= \tau_1 - \tau_{1'}, \, {\rho} = r_1 - r_{1'}\non\\
	\partial_{\tau_1} &= \partial_t + \frac{1}{2} \partial_T, \, \partial_{\tau_{1'}} = -\partial_t + \frac{1}{2} \partial_T.
\end{align}
Then the LHS and RHS of Eq.~\eqref{appeq:Dyson0} read
\begin{align}
	{\rm LHS} &= (-\partial_t - \frac{1}{2} \partial_T + \mu) \mG(R, T; \rho, t) - \left( \sum_{\delta} (- {\msf{t}_{\delta}} )\eu^{\iu \ch {A}_{R + \rho/2 - \delta/2}\cdot  \delta}\right)\mG(R-\delta/2, T; \rho-\delta, t)\non\\
	{\rm RHS} & = \int \ddiff{3} \mSigma(R+r_3/2, T+\tau_3/2; \rho- r_3, t-\tau_3) \mG(R + r_3/2 - \rho/2, T + \tau_3/2 - t/2; r_3, \tau_3) + \delta (\rho) \delta(t).
\end{align}

Define the Fourier transform for the relative coordinates of a bilocal field $\mathcal{F}(R, T; \rho, t)$ as
\begin{align}
	\mathcal{F}(R, T; \rho, t)
	=\eu^{\iu \ch {A}_{R}\cdot \rho}\, \mathcal{F}'(R, T; \rho, t)
	= \frac{1}{\beta}\sum_{n} \int \frac{\diff^2 p}{(2\pi)^2}
	\eu^{\iu (p + \ch A_R)\cdot \rho- \iu \omega_n t}\,
	\mathcal{F}'(R, T; p, \omega_n).
	\label{appeq:FourierBilocal}
\end{align}

Using $\mG(R-\delta/2, T; \rho-\delta, t)=\eu^{-\frac{\delta}{2}\cdot \nabla_R} \mG(R, T; \rho-\delta, t)$ and $\ch A_R \cdot \rho = \frac{\theta_{ab}}{2} R_a \rho_b$ in the symmetric gauge, we obtain
\begin{align}
	{\rm LHS} & = \beta^{-1} \sum_{n} \int  \frac{\diff^2 p}{(2\pi)^2} \,  \eu^{\iu p\cdot \rho- \iu \omega_n t} \non\\
	& \quad\quad\quad\left(( \iu \omega_n -\frac{1}{2} \partial_T+ \mu)\eu^{\iu \ch {A}_{R}\cdot \rho}- \sum_\delta(- {\msf{t}_{\delta}} ) \eu^{\iu \frac{ \theta_{ab}}{2} (R+\rho/2-\delta/2)_a \delta_b} \eu^{-\iu p \cdot \delta} \eu^{-\frac{\delta}{2} \cdot \nabla_R}\eu^{\iu \frac{\theta_{ab}}{2}R_a (\rho-\delta)_b} \right) \mG'(R, T; p, \omega_n)\non\\
	& = \beta^{-1}\sum_{n} \int  \frac{\diff^2 p}{(2\pi)^2} \,  \eu^{\iu (p+\ch A_R)\cdot \rho- \iu \omega_n t} \left(( \iu \omega_n -\frac{1}{2} \partial_T+ \mu)- \sum_\delta(- {\msf{t}_{\delta}} ) \eu^{-\iu p \cdot \delta} \eu^{\iu \frac{ \theta_{ab}}{2} \rho_a \delta_b}  \eu^{-\frac{\delta}{2} \cdot \nabla_R} \right) \mG'(R, T; p, \omega_n)\non\\
	& = \beta^{-1}\sum_{n} \int  \frac{\diff^2 p}{(2\pi)^2} \,  \eu^{\iu (p + \ch A_R)\cdot \rho- \iu \omega_n t}\left(( \iu \omega_n -\frac{1}{2} \partial_T+ \mu) \mG'(R, T; p, \omega_n)- \mathsf{h}_p \left(\eu^{\frac{\iu \theta_{ab}}{2}\overleftarrow{\partial_a}\overrightarrow{\partial_b}} \right) \eu^{-\frac{\delta}{2} \cdot \nabla_R} \mG'(R, T; p, \omega_n) \right) 
	\label{appeq:LHS}
\end{align}
From the second to the third line, we used
$$\eu^{-\frac{\delta}{2} \cdot \nabla_R}\eu^{\iu \frac{\theta_{ab}}{2}R_a (\rho-\delta)_b} = \eu^{\iu \frac{\theta_{ab}}{2}R_a (\rho-\delta)_b} \eu^{-\frac{\delta}{2} \cdot \nabla_R} \eu^{-\iu \frac{\theta_{ab}}{4} \delta_a \rho_b}. $$
From the third to the last line, integration by parts was applied, and the left (right) arrows in $\eu^{\frac{\iu \theta_{ab}}{2}\overleftarrow{\partial_a}\overrightarrow{\partial_b}}$ act on $\mathsf{h}_p$ and $\mG$, respectively. Equation~\eqref{appeq:LHS} is invariant under translations of the CoM coordinates, i.e.\ $R\rightarrow R+\Delta R$ and $T\rightarrow T+\Delta T$. Therefore, the dependence on $R,T$ can be dropped, and we write $\mG'(p, \omega_n) = \eu^{-\frac{\delta}{2} \cdot \nabla_R} \mG'(R, T; p, \omega_n)$.

Similarly,
\begin{align}
	{\rm RHS} & = \int \diff r_3 \diff \tau_3 \, \mSigma(R+r_3/2, T+\tau_3/2; \rho- r_3, t-\tau_3) \mG(R + r_3/2 - \rho/2, T + \tau_3/2 - t/2; r_3, \tau_3) + \delta(\rho) \delta(t)\non\\
	& = \delta(\rho) \delta(t)  + \int \diff r_3 \diff \tau_3 \, \beta^{-1}\sum_n \int  \frac{\diff^2 p}{(2\pi)^2}  \eu^{\iu (p +\ch A_{R+r_3/2})\cdot (\rho-r_3) - \iu \omega_n (t-\tau_3)}\mSigma'(R+r_3/2, T+\tau_3/2; p, \omega_n) \non\\
	& \qquad \qquad \qquad \qquad \beta^{-1}\sum_{n'} \int  \frac{\diff^2 p'}{(2\pi)^2}  \eu^{\iu (p'+\ch A_{R+r_3/2-\rho/2})\cdot r_3 - \iu \omega_{n'} \tau_3} \mG'(R + r_3/2 - \rho/2, T + \tau_3/2 - t/2;p', \omega_{n'}) \non\\
	& = \delta(\rho) \delta(t)+\beta^{-1}\sum_n \eu^{-\iu \omega_n t} \int  \frac{\diff^2 p}{(2\pi)^2}   \frac{\diff^2 p'}{(2\pi)^2}  \int \diff r_3 \, \eu^{\iu (p+\ch A_R) \cdot \rho} \eu^{\iu (p'-p)\cdot r_3} \eu^{\frac{\iu \theta_{ab}}{2}(r_3)_a \rho_b}  \mSigma'(p,\omega_n) \mG'(p', \omega_{n}) \non\\
	& = \beta^{-1}\sum_n \int  \frac{\diff^2 p}{(2\pi)^2}   \eu^{\iu (p+\ch A_R) \cdot \rho -\iu \omega_n t} \left( \mSigma'(p,\omega_n) \left(\eu^{\frac{\iu \theta_{ab}}{2}\overleftarrow{\partial_a}\overrightarrow{\partial_b}} \right) \mG'(p, \omega_{n}) + 1\right)
	\label{appeq:RHS}
\end{align}
From the second to the third equality, we used the translational invariance of $\mG'$ and $\mSigma'$.

Equating Eqs.~\eqref{appeq:LHS} and~\eqref{appeq:RHS}, we obtain
\begin{align}
	\left(( \iu \omega_n -\frac{1}{2} \partial_T+ \mu) - \left(\mathsf{h}_p + \mSigma'(p,\omega_n) \right) \eu^{\frac{\iu \theta_{ab}}{2}\overleftarrow{\partial_a}\overrightarrow{\partial_b}} \right)  \mG'( p, \omega_n) = 1.
	\label{appeq:DysonB}
\end{align}

\section{Review of Luttinger--Ward functional}
\label{app:LWintroduction}
The Luttinger--Ward functional provides a representation of the grand potential in terms of the two-point Green's function, assuming the skeleton (2PI) expansion is convergent (or at least asymptotic). Here we review the statement and outline a proof.

\emph{Statement:} Given interactions, one can define a functional $\tilde{\Omega}_{\rm int}(\tilde{\mG}, \tilde{\mSigma})$,
\begin{align}
	\tilde{\Omega}_{\rm int}(\tilde{\mG}, \tilde{\mSigma})
	=\tilde{\Phi}(\tilde{\mG})- T \Tr \ln \tilde{\mG}^{-1} - T \Tr \tilde{\mSigma} \circ \tilde{\mG},
\end{align}
where $\tilde{\Phi}(\tilde{\mG})$ is the sum of all 2PI diagrams. The physical grand potential $\Omega_{\rm int}$ is
\begin{align}
	\Omega_{\rm int}=\tilde{\Omega}_{\rm int}(\mG, \mSigma)=\Phi(\mG)- T \Tr \ln \mG^{-1} - T \Tr \mSigma \circ \mG.
	\label{appeq:LWfunction}
\end{align}
The two-point functions $\mG_{12}$ and $\mSigma_{12}$ satisfy the coupled equations
\begin{align}
	\begin{cases}
		\text{Dyson equation:} & \left( \mG_0^{-1}-\mSigma\right)_{12}={\mG^{-1}}_{12} \ \leftrightarrow\ \left(\left( \mG_0^{-1}-\mSigma\right)\circ \mG \right)_{1 1'}=\delta_{1 1'}, \\
		\text{Self-energy from 2PI functional:} & \frac{\delta \Phi(\mG)}{\delta \mG_{12}}=T\,\mSigma_{21}.
	\end{cases}
	\label{appeq:LWconstraints}
\end{align}
In the Dyson equation we used $({\mG^{-1}}\circ \mG)_{11'}=\delta_{1 1'}$, where the bilocal convolution is
\begin{equation}
	(f\circ g )_{1 1'} =\int \ddiff{2} f_{12} \, g_{2 1'}, \qquad
	\Tr(f \circ g) = \tr\left( \int \ddiff{1} \ddiff{1'} \delta_{11'} (f \circ g)_{11'} \right).
\end{equation}
Here, $\Tr$ sums over internal and space-time coordinates [$l=1,2,...$ for $(\tau_l, r_l)$], $\int \ddiff{1} = \int \diff \tau_1 \diff^2 r_1$, and $\tr$ denotes the trace over internal indices only.

\emph{Proof (outline):} The grand potential satisfies $\frac{\delta \Omega_{\rm int}}{\delta {\mG_0^{-1}}\,_{12}}=- T \mG_{21}$, which follows directly from the path-integral definition. Together with the Dyson equation in \eqref{appeq:LWconstraints}, this implies that $\Phi(\mG)$ generates the self-energy through functional differentiation. The main steps are:\\
{[1]} The identity $\frac{\delta \tilde{\Omega}_{\rm int}}{\delta {{\tilde{\mG}_0}^{-1}}{}_{12}}=- T \tilde{\mG}_{21}$ holds at the functional level for any trial $\tilde{\mG}_0^{-1}$. \\
{[2]} Performing a Legendre transform yields the ``dual'' functional $\tilde{F}_{\rm int} (\tilde{\mG}) = \tilde{\Omega}_{\rm int}(\tilde{\mG}_0^{-1}) + T \Tr \tilde{\mG}_{0}^{-1} \circ \tilde{\mG}$, with
\begin{align}
	\frac{\delta \tilde{F}_{\rm int}}{\delta {\tilde{\mG}}_{12}} &= T {{\tilde{\mG}_0}^{-1}}\,_{21}.
\end{align}
{[3]} Define $\tilde{\Phi} = \tilde{\Omega}_{\rm int} + T \Tr \tilde{\mG}_{0}^{-1}\circ \tilde{\mG} + T \Tr \ln \tilde{\mG}^{-1}$, such that
\begin{align}
	\frac{\delta \tilde{\Phi}}{\delta {\tilde{\mG}}_{12}}
	= T {{\tilde{\mG}_0}^{-1}}{}_{21} - T {{\tilde{\mG}}^{-1}}_{21}.
	\label{eq:PhiFunc}
\end{align}
Here, $\frac{\delta \Tr \ln \tilde{\mG}^{-1}}{\delta \tilde{\mG}_{12}}=-{{\tilde{\mG}}^{-1}}_{21}$ follows from standard matrix identities. \\
{[4]} At the physical point $\tilde{\mG}\to \mG$ and $\tilde{\mG}_0\to \mG_0$, the Dyson equation implies $\frac{\delta \Phi}{\delta \mG_{12}} = T \mSigma_{21}$, i.e.\ the second equation in \eqref{appeq:LWconstraints}. Thus $\Phi$ is the sum of 2PI skeleton diagrams. \\
{[5]} For a given bare kernel $\mG_0^{-1}$ (with or without external fields), solving \eqref{appeq:LWconstraints} self-consistently determines $\mG$ and $\mSigma$, and the grand potential follows from \eqref{appeq:LWfunction}.

\section{Generalized Fourier transformation}
\label{app:Fourier}
In this appendix we derive Eqs.~\eqref{eq:OmegaA1} and~\eqref{eq:OmegaA2} from Eq.~\eqref{eq:OmegaA0} using the generalized Fourier transformation defined in Eq.~\eqref{appeq:generalizedF}. We first discuss the generalized Fourier transformation and the motivation for introducing it in App.~\ref{appsec:FourierA}. We then apply it to derive Eq.~\eqref{eq:OmegaA1} in App.~\ref{appsec:FourierB}. Finally, we show that the action can be written in terms of commuting coordinates equipped with the Moyal algebra, and derive Eq.~\eqref{eq:OmegaA2} in App.~\ref{appsec:FourierC}. 

\subsection{Basics on the generalized Fourier transformation}
\label{appsec:FourierA}
To simplify notation, we define the noncommutativity scale $\theta=\ell_B^{-2}$, so that
\begin{align}
	[\hat{p}_a, \hat{p}_b] = \iu \epsilon_{ab} \theta = \iu \theta_{ab}.
\end{align}

The generalized Fourier transform is obtained by replacing the plane wave $\eu^{\iu p \cdot r}$ with the operator-valued factor $\eu^{\iu \hat{p}\cdot r}$. To work with c-numbers rather than operators, it is convenient to introduce a coherent-state basis,
\begin{align}
	\cket{\eta} &= \eu^{{\eta} \hat{b}^\dg / \sqrt{2\theta}}\ket{0}, \, \quad \hat{b} \cket{\eta} = {\eta}/\sqrt{2\theta} \cket{\eta} \non\\
	\cbra{\xi} &= \bra{0} \eu^{\bar{\xi} \hat{b} / \sqrt{2\theta}}, \, \quad \cbra{\xi} \hat{b}^\dg = \cbra{\xi} \bar{\xi}/\sqrt{2\theta},
\end{align}
where $\hat{b}, \hat{b}^\dg$ are defined as
\begin{align}
	\hat{b} = \frac{\hat{p}_x + \iu \hat{p}_y}{\sqrt{2\theta}}, \quad \hat{b}^\dg = \frac{\hat{p}_x - \iu \hat{p}_y}{\sqrt{2\theta}}.
\end{align}

The generalized Fourier transform then reads
\begin{align}
	{O}_r = \cTr \left(\eu^{\iu \hat{p} r}{O}_{\hat{p}}\right), \qquad
	{O}_{\hat{p}} = \int \diff^2 r \eu^{-\iu \hat{p} r} {O}_r.
	\label{appeq:generalizedF}
\end{align}
Here $\cTr\left(\cdot\right)=\int \frac{\diff^2 \xi}{(2\pi)^2} \eu^{-\abs{\xi}^2/(2\theta)} \cbra{\xi} \cdot \cket{\xi}$ with $\diff^2 \xi = \frac{1}{2} \diff \bar{\xi} \diff \xi$, and $\diff^2 r = \diff r_x \diff r_y = \frac{1}{2}\diff \bar{z}\diff z$, where $z=r_x + \iu r_y$ and $\bar{z}=r_x - \iu r_y$.

To establish Eq.~\eqref{eq:generalizedF}, we used the coherent-state identities
\begin{align}
	\cbra{\xi} \eu^{\iu \hat{p} \cdot r }\cket{\eta} &= \eu^{\frac{\iu}{2} (\bar{\xi} z + \bar{z}\eta)} \eu^{-\frac{\theta}{4} \bar{z}z} \eu^{\bar{\xi}\eta/2\theta}\non\\
	\cbra{\xi} \eu^{\iu \hat{p} \cdot r }\eu^{\iu \hat{p} \cdot r' }\cket{\eta} &= \cbra{\xi} \eu^{\iu \hat{p} \cdot (r+r') }\cket{\eta} \eu^{-\frac{\theta}{4} (\bar{z}z'-z\bar{z}')}.
\end{align}
The completeness relation reads $\cbra{\xi} \eta ) = \eu^{\bar{\xi} {\eta}/2\theta}$ and $\mathbb{1} = \int \diff (\xi, \bar{\xi}) \eu^{-\bar{\xi}\xi/2\theta} \cket{\xi} \cbra{\xi}$, with $\diff(\xi, \bar{\xi})= \frac{\diff \xi \diff \bar{\xi}}{4\pi \theta}$.

\subsection{Derivation of Eq.~\eqref{eq:OmegaA1}}
\label{appsec:FourierB}
The action in Eq.~\eqref{eq:OmegaA0} in Matsubara frequency reads
\begin{align}
	\mathcal{S}_E &= \sum_{\omega_n} \int \diff^2 r \diff^2 r' \bar{\psi}_n(r) \left((-\iu \omega_n - \mu) \delta(r-r') +\sum_{\delta}  (- {\mathsf{t}_{\delta}}) \eu^{\iu \ch {A}_{(r + r' )/2} \cdot {\delta}}\delta (r-\delta-r')  + \mSigma(r, r'; \omega_n) \right) \psi_n(r')\non\\
	& = \sum_{\omega_n} \int \diff^2 r \diff^2 r' \bar{\psi}_n(r) \eu^{\iu \ch {A}_{(r + r' )/2} \cdot (r-r')} \left((-\iu \omega_n - \mu) \delta(r-r') +  \left(\sum_{\delta}  (- {\mathsf{t}_{\delta}})\delta (r-r'-\delta)  + \mSigma'(r-r'; \omega_n)\right) \right) \psi_n(r').
\end{align}
Here $\psi (\tau) = \frac{1}{\sqrt{\beta}} \sum_n \eu^{-\iu \omega_n \tau} \psi_n$ and $\mSigma(r, \tau; r', \tau') = \frac{1}{\beta}\sum_n \eu^{-\iu \omega_n (\tau - \tau')} \mSigma(r,r'; \omega_n)$. From the Dyson equation, $\mSigma$ is time-translation invariant, and $\mSigma'$ is space-translation invariant. From the first to the second line, we used $\mSigma(r, r'; \omega_n) = \eu^{\iu \ch A_{(r+r')/2}\cdot (r-r')} \mSigma'(r-r'; \omega_n)$. The term enclosed by the outermost bracket in the second line is translationally invariant, which we denote by
\begin{align}
	g^{-1}(r-r') = (-\iu \omega_n - \mu) \delta(r-r') +  \left(\sum_{\delta}  (- {\mathsf{t}_{\delta}})\delta (r-r'-\delta)  + \mSigma'(r-r'; \omega_n)\right).
\end{align}

Next, we perform the generalized Fourier transformation on $\bar{\psi}$ and $\psi$:
\begin{align}
	\mathcal{S}_E & = \int \diff^2{r} \diff^2{r'}  \cTr\left({\bar{\psi}}_{\hat{p}} \eu^{-\iu \hat{p} \cdot r}\right) \eu^{\iu \ch {A}_{(r + r' )/2} \cdot (r-r')} g^{-1}(r-r') \cTr\left(\eu^{\iu \hat{p} \cdot r'}\psi_{\hat{p}}  \right)\non\\ 
	& = \int \diff^2{r} \diff^2{\rho}  \cTr\left({\bar{\psi}}_{\hat{p}} \eu^{-\iu \hat{p} \cdot r}\right) \eu^{\iu \frac{\theta_{ab}}{2} r_a \rho_b} g^{-1}(\rho) \cTr\left( \eu^{\iu \hat{p} \cdot (r-\rho)} \psi_{\hat{p}}  \right)\non\\
	& = (\frac{\theta}{2\pi})^2 \,\int \frac{\diff \bar{z} \diff z}{2} \frac{\diff \bar{\varrho} \diff \varrho}{2} \frac{\diff \bar{\chi} \diff \chi}{4\pi \theta}  \frac{\diff \bar{\nu} \diff \nu}{4\pi \theta}  \frac{\diff \bar{\xi} \diff \xi}{4\pi \theta} \frac{\diff \bar{\eta} \diff \eta}{4\pi \theta} \, g^{-1}(\rho) \eu^{\frac{\theta}{4} (\bar{z} \varrho - z \bar{\varrho})} \non\\
	&  \qquad \qquad \qquad \qquad \qquad
	\eu^{-\bar{\chi}\chi/2\theta - \bar{\nu}\nu /2\theta}\cbra{\chi} \bar{\psi}_{\hat{p}} \cket{\nu} \cbra{\nu} \eu^{-\iu \hat{p} \cdot r} \cket{\chi}  \eu^{-\bar{\xi}\xi/2\theta - \bar{\eta}\eta /2\theta}\cbra{\xi} \eu^{\iu \hat{p} \cdot (r-\rho)}  \cket{\eta} \cbra{\eta}  {\psi}_{\hat{p}} \cket{\xi}  \non\\
	& = (\frac{\theta}{2\pi})^2 \,\int \frac{\diff \bar{z} \diff z}{2} \frac{\diff \bar{\varrho} \diff \varrho}{2} \frac{\diff \bar{\chi} \diff \chi}{4\pi \theta}  \frac{\diff \bar{\nu} \diff \nu}{4\pi \theta}  \frac{\diff \bar{\xi} \diff \xi}{4\pi \theta} \frac{\diff \bar{\eta} \diff \eta}{4\pi \theta} \, g^{-1}(\rho) \eu^{\frac{\theta}{4} (\bar{z} \varrho - z \bar{\varrho})} \non\\
	&  \qquad  \qquad
	\eu^{-\bar{\chi}\chi/2\theta - \bar{\nu}\nu /2\theta}\cbra{\chi} \bar{\psi}_{\hat{p}} \cket{\nu} \eu^{-\iu \frac{1}{2} (\bar{\nu}z+ \chi \bar{z}) - \frac{\theta}{4} \bar{z}z + \bar{\nu}\chi/2\theta} \eu^{-\bar{\xi}\xi/2\theta - \bar{\eta}\eta /2\theta} \eu^{\iu \frac{1}{2} (\bar{\xi}(z-\varrho)+ \eta (\bar{z}-\bar{\varrho})) - \frac{\theta}{4} (\bar{z}-\bar{\varrho})(z-\varrho) + \bar{\xi}\eta/2\theta}  \cbra{\eta}  {\psi}_{\hat{p}} \cket{\xi}  \non\\
	& =  (\frac{\theta}{2\pi})^2 \, \int \frac{\diff \bar{\varrho} \diff \varrho}{2} \frac{\diff \bar{\chi} \diff \chi}{4\pi \theta}  \frac{\diff \bar{\nu} \diff \nu}{4\pi \theta}  \frac{\diff \bar{\xi} \diff \xi}{4\pi \theta} \frac{\diff \bar{\eta} \diff \eta}{4\pi \theta} \, g^{-1}(\rho)  \eu^{-\bar{\chi}\chi/2\theta - \bar{\nu}\nu /2\theta-\bar{\xi}\xi/2\theta - \bar{\eta}\eta /2\theta + \bar{\nu}\chi/2\theta + \bar{\xi}\eta/2\theta}  \cbra{\chi} \bar{\psi}_{\hat{p}} \cket{\nu} \cbra{\eta}  {\psi}_{\hat{p}} \cket{\xi}  \non\\
	&  \qquad \qquad \int \frac{\diff \bar{z} \diff z}{2} \, \eu^{\frac{\theta}{4} (\bar{z} \varrho - z \bar{\varrho})}  \eu^{-\iu \frac{1}{2} (\bar{\nu}z+ \chi \bar{z}) - \frac{\theta}{4} \bar{z}z }\eu^{\iu \frac{1}{2} (\bar{\xi}(z-\varrho)+ \eta (\bar{z}-\bar{\varrho})) - \frac{\theta}{4} (\bar{z}-\bar{\varrho})(z-\varrho) }   \non\\
	& =  \frac{\theta}{2\pi}\int \frac{\diff \bar{\varrho} \diff \varrho}{2} \frac{\diff \bar{\chi} \diff \chi}{4\pi \theta}  \frac{\diff \bar{\nu} \diff \nu}{4\pi \theta}  \frac{\diff \bar{\xi} \diff \xi}{4\pi \theta} \frac{\diff \bar{\eta} \diff \eta}{4\pi \theta} \, g^{-1}(\rho)  \eu^{-\bar{\chi}\chi/2\theta - \bar{\nu}\nu /2\theta-\bar{\xi}\xi/2\theta - \bar{\eta}\eta /2\theta + \bar{\nu}\chi/2\theta + \bar{\xi}\eta/2\theta}  \cbra{\chi} \bar{\psi}_{\hat{p}} \cket{\nu} \cbra{\eta}  {\psi}_{\hat{p}} \cket{\xi}  \non\\
	&  \qquad \qquad \eu^{-\iu \frac{1}{2}(\bar{\xi}\varrho+\eta \bar{\varrho})} \eu^{-\frac{\theta}{4} \bar{\varrho}\varrho} \eu^{-(-\bar{\nu}+\bar{\xi})(-\chi + \eta)/2\theta} \eu^{\iu \frac{1}{2}(-\bar{\nu}+\bar{\xi})\varrho}\non\\
	& =  \frac{\theta}{2\pi} \int \frac{\diff \bar{\varrho} \diff \varrho}{2} \frac{\diff \bar{\chi} \diff \nu}{4\pi \theta}  \frac{\diff \bar{\nu} \diff \eta}{4\pi \theta}  \frac{\diff \bar{\eta} \diff \xi}{4\pi \theta}  \, g^{-1}(\rho)  \eu^{ - \bar{\nu}\nu /2\theta - \bar{\eta}\eta /2\theta }  \eu^{-\iu \frac{1}{2}(\bar{\nu}\varrho+\eta \bar{\varrho})} \eu^{-\frac{\theta}{4} \bar{\varrho}\varrho} \eu^{\bar{\nu}\eta/2\theta} \cbra{\chi} \bar{\psi}_{\hat{p}} \cket{\nu} \cbra{\eta}  {\psi}_{\hat{p}} \cket{\xi} \non\\
	& \qquad \qquad \int \frac{\diff \bar{\xi} \diff \chi}{4\pi \theta} \,  \eu^{(\bar{\xi}-\bar{\chi})(\chi - \xi)/2\theta} \eu^{-\bar{\chi}\xi/2\theta}\non\\
	& =  \frac{\theta}{2\pi} \int \frac{\diff \bar{\varrho} \diff \varrho}{2} \frac{\diff \bar{\chi} \diff \nu}{4\pi \theta}  \frac{\diff \bar{\nu} \diff \eta}{4\pi \theta}  \frac{\diff \bar{\eta} \diff \xi}{4\pi \theta}  \, \cbra{\chi} \bar{\psi}_{\hat{p}} \cket{\nu} \eu^{ - \bar{\nu}\nu /2\theta}   g^{-1}(\rho)   \eu^{-\iu \frac{1}{2}(\bar{\nu}\varrho+\eta \bar{\varrho})} \eu^{-\frac{\theta}{4} \bar{\varrho}\varrho} \eu^{\bar{\nu}\eta/2\theta}  \eu^{ - \bar{\eta}\eta /2\theta } \cbra{\eta}  {\psi}_{\hat{p}} \cket{\xi}  \eu^{-\bar{\chi}\xi/2\theta}\non\\
	& =   \int \frac{\diff \bar{\varrho} \diff \varrho}{2} \frac{\diff \bar{\chi} \diff \xi}{ 8 \pi^2}  \frac{\diff \bar{\nu} \diff \nu}{4\pi \theta}  \frac{\diff \bar{\eta} \diff \eta}{4\pi \theta}  \, \cbra{\chi} \bar{\psi}_{\hat{p}} \cket{\nu} \eu^{ - \bar{\nu}\nu /2\theta}   g^{-1}(\rho)   \cbra{\nu} \eu^{-\iu \hat{p}\cdot \rho} \cket{\eta}  \eu^{ - \bar{\eta}\eta /2\theta } \cbra{\eta}  {\psi}_{\hat{p}} \cket{\xi}  \eu^{-\bar{\chi}\xi/2\theta}\non\\
	& = \cTr\left(\bar{\psi}_{\hat{p}}   (g^{-1})_{\hat{p}}   {\psi}_{\hat{p}}\right),
\end{align}
where $(g^{-1})_{\hat{p}} = \int \diff^2 \rho\, g^{-1}(\rho) \eu^{-\iu \hat{p}\cdot \rho}$.

\subsection{Derivation of Eq.~\eqref{eq:OmegaA2}}
\label{appsec:FourierC}
The fields $\bar{\psi}, \psi$ and the kernel $g^{-1}$ are defined in the noncommuting coordinates $\hat{p}$. Since the commutator is a constant, $[\hat{p}_a, \hat{p}_b] = \iu \theta \epsilon_{ab}$, the action can be rewritten in terms of commuting variables, with the operator algebra replaced by the Moyal algebra. Using~\cite{Szabo2003,Douglas2001rmp}
\begin{align}
	\cTr\left(\mathcal{A}_{\hat{p}}\mathcal{B}_{\hat{p}}\mathcal{C}_{\hat{p}} ...  \right) = \int \frac{\diff^2 p}{(2\pi)^2} \mathcal{A}(p)\star \mathcal{B}(p) \star \mathcal{C}(p) ...
\end{align}
where
\begin{align}
	\mathcal{A}_p = \int \diff^2 r\, \mathcal{A}(r)\eu^{-\iu p \cdot r},
\end{align}
we obtain
\begin{align}
	\mathcal{S}_E = \int \frac{\diff^2 p}{(2\pi)^2} \bar{\psi}_p \star (g^{-1})_p \star \psi_p,
\end{align}
where
\begin{align}
	(g^{-1})_p = \int \diff^2 \rho\, g^{-1}(\rho) \eu^{-\iu p \cdot \rho} = - \iu \omega_n - \mu + \left(\mathsf{h}_p + \mSigma'(p,\omega_n)\right),
\end{align}
with $\mathsf{h}_p = \sum_{\delta} (-\mathsf{t}_{\delta}) \eu^{-\iu p \cdot \delta}$.

\end{document}